\documentclass[sigconf,nonacm]{acmart}

\usepackage{booktabs}
\usepackage{array}
\usepackage{enumitem}
\usepackage{multirow}
\usepackage{xurl}
\usepackage{xcolor}
\usepackage{graphicx}

\definecolor{scenarioeducation}{HTML}{3F7CAC}
\definecolor{scenarioworkplace}{HTML}{6B7F3E}
\definecolor{scenariocreative}{HTML}{9A5C7B}
\definecolor{scenariohealthcare}{HTML}{32746D}
\definecolor{scenariopolitics}{HTML}{B86E31}
\definecolor{scenariointerpersonal}{HTML}{7D6AA8}
\definecolor{artifacttask}{HTML}{4B5563}

\newcommand{\colorlabel}[2]{%
  \textcolor{#1}{\raisebox{0.08ex}{\rule{0.62em}{0.62em}}\hspace{0.35em}\textbf{#2}}}
\newcommand{\scenariolabel}[2]{\colorlabel{#1}{#2}}
\newcommand{\tasklabel}[1]{\colorlabel{artifacttask}{#1}}

\newif\ifshownotes
\shownotestrue

\title{Who Bears the Cost of Honesty?\\
A FAccT Workshop Synthesis and Research Agenda for Equitable AI Disclosure}

\author{Runlong Ye}
\affiliation{%
  \institution{University of Toronto}
  \city{Toronto}
  \state{Ontario}
  \country{Canada}}
\email{harryye@cs.toronto.edu}

\author{Jessica He}
\affiliation{%
  \institution{IBM Research}
  \city{Cambridge}
  \state{Massachusetts}
  \country{United States}}
\email{jessicahe@ibm.com}

\author{Finola Finn}
\affiliation{%
  \institution{University of Luxembourg}
  \city{Esch-sur-Alzette}
  \country{Luxembourg}}
\email{finola.finn@uni.lu}

\author{Angel Hsing-Chi Hwang}
\affiliation{%
  \institution{University of Southern California}
  \city{Los Angeles}
  \state{California}
  \country{United States}}
\email{angel.hwang@usc.edu}

\author{Donal Khosrowi}
\affiliation{%
  \institution{Leibniz University Hannover}
  \city{Hannover}
  \country{Germany}}
\email{donal.khosrowi@philos.uni-hannover.de}

\author{Seyun Kim}
\affiliation{%
  \institution{Carnegie Mellon University}
  \city{Pittsburgh}
  \state{Pennsylvania}
  \country{United States}}
\email{seyunkim@cs.cmu.edu}

\author{Morgan Klaus Scheuerman}
\affiliation{%
  \institution{Sony AI}
  \city{Barcelona}
  \country{Spain}}
\email{morgan.scheuerman@sony.com}

\begin{document}

\begin{teaserfigure}
  \centering
  \includegraphics[width=.7\textwidth]{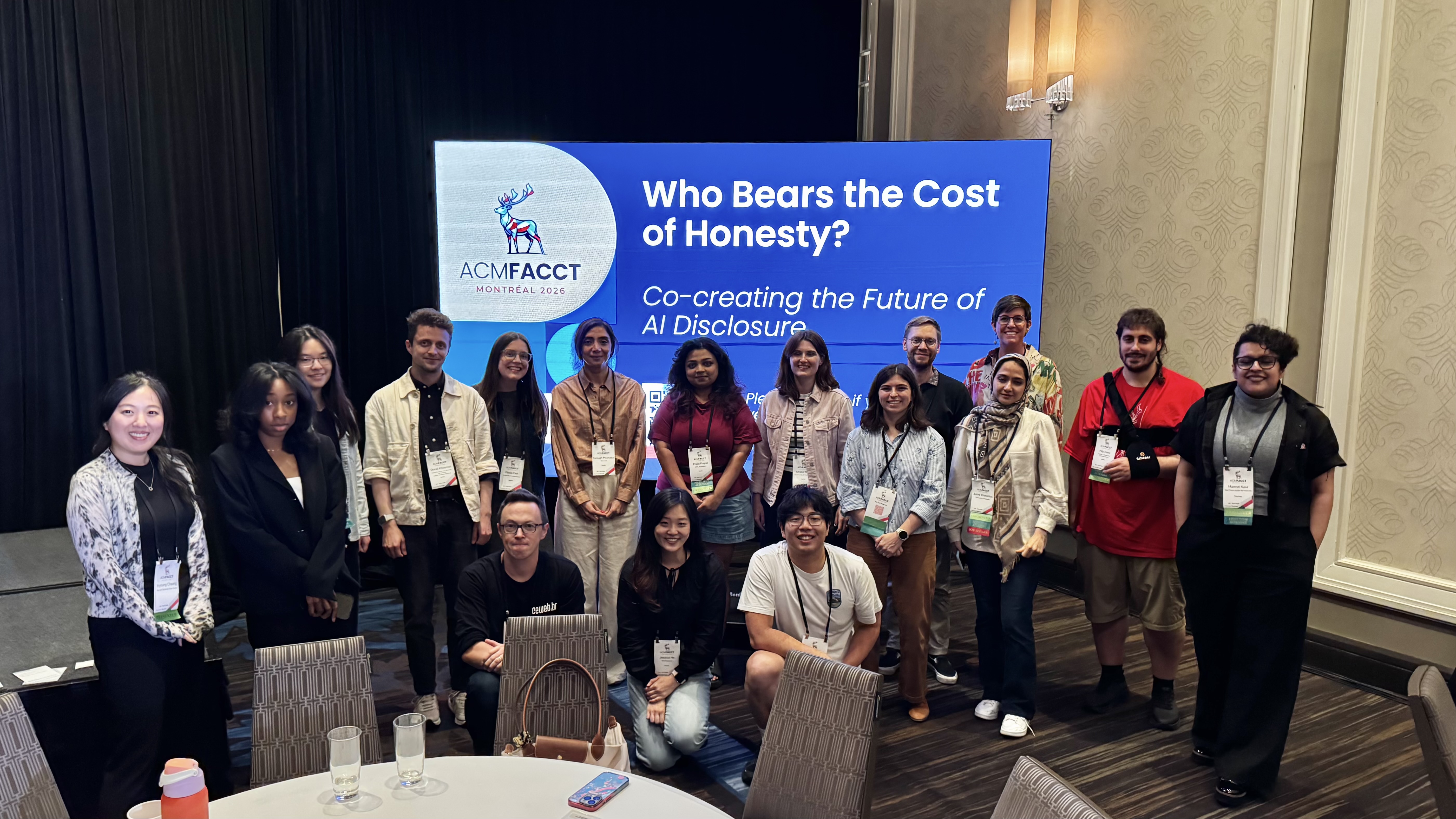}
  \caption{The FAccT 2026 \textit{Who Bears the Cost of Honesty?} Workshop was held in Montreal, Canada, on June 26, 2026.}
  \label{fig:teaser}
  \Description{Group photo for FAccT workshop}
\end{teaserfigure}

\begin{abstract}
AI disclosure is increasingly promoted and sometimes required as a route to transparency, accountability, provenance, and trust. Yet disclosure can also expose AI users to suspicion, stigma (e.g., competence penalties), and surveillance, affecting minoritized groups in particular. This paper reports on \emph{Who Bears the Cost of Honesty?}, a CRAFT workshop at the 2026 ACM Conference on Fairness, Accountability, and Transparency that used scenario-anchored power mapping and design fiction to explore the benefits, harms, tensions, and power asymmetries that emerge under AI disclosure norms and mandates. We document the workshop design and analyze the disclosure approaches participants co-created, comprising four completed power maps, three context cards, and one interface prototype. These artifacts span education, workplace, politics/journalism, and interpersonal contexts. They depict disclosure as a multi-actor accountability process, surface concerns that the use of accessibility-related AI could be held against workers in performance evaluations, and explore how context-specific, bottom-up disclosures may support transparency while mitigating some risks of stigma and misinterpretation. We contribute (1) a documented two-stage workshop method; (2) an artifact-grounded thematic synthesis; and (3) a diagnostic framework, the \emph{Cost-of-Honesty Stack}, with provisional design suggestions and research directions.
\end{abstract}


\maketitle

\section{Introduction}

AI disclosure is becoming a common response to uncertainty around generative AI usage---if AI was involved, say so. Such disclosure can help establish provenance, reduce deception, support accountability, and calibrate trust. Yet revealing AI use is not a neutral act. In workplace evaluations, AI users can face competence and motivation penalties \citep{reif2025social}, and even suspected AI use can lower evaluations of writers and willingness to hire them \citep{kadoma2025perceptual}. Evidence also suggests that these costs disproportionately affect minoritized users, such as female and mature-age engineers \citep{gai2025competence}. Disclosure requirements are often justified by what recipients, institutions, or the public may \emph{gain}---including trust, consent, provenance, evaluative confidence, or liability protection---yet less attention is paid to what the person disclosing may \emph{surrender}, including privacy, credit, autonomy, reputation, or the ability to use assistance without revealing a disability, language background, or precarious working condition. We call this distributional question the \emph{cost of honesty}: who receives disclosure's benefits, who bears its burdens, and who has the power to decide whether those burdens are proportionate?

Understanding the cost of honesty requires clarity about what disclosure is intended to accomplish. Wittenberg et al. distinguish process-based labels from labels intended to reduce deceptive effects \citep{wittenberg2024labeling}. Human-centered work likewise shows that even a legal obligation leaves open who must disclose what, to whom, when, and how \citep{elali2024transparent}. Institutions are acting before these questions are settled. Article~50 of the EU AI Act and its implementation code establish transparency expectations for certain AI systems and content, while ACM policy requires authors to disclose generative-AI-created content and remain accountable for its accuracy and integrity \citep{europeanunion2024aiact,europeancommission2026transparencycode,acm2025authorship}. But neither the stated purpose of disclosure nor the existence of a formal rule determines how disclosure will operate in practice. Transparency is not only a property of a label, but a process that operates through institutions, roles, incentives, and relationships \citep{ananny2018seeing,selbst2019fairness}. Coursework submission forms, performance reviews, newsroom notes, and dating-app controls may ask similar questions while giving different actors the power to demand, retain, interpret, and act on the answers.

As a result, consequences and disclosure expectations remain inconsistent within and across domains. In education, rewriting and translation can support non-native English writers, while unreliable detectors can disproportionately flag their work \citep{ito2023rewriting,liang2023detectors}. Among computer science students, preferred disclosure requirements varied with the level and type of AI assistance, while personal intentions to disclose did not follow the same authorship and grading-fairness judgments \citep{ye2026norms}. Elsewhere, disclosure expectations remain contested in freelance work \citep{hwang2026freelance}. Credit judgments vary with the type, amount, and initiative of human and AI contributions \citep{he2025credit}, and human--AI co-writing raises distinct concerns about voice and authenticity \citep{hwang2025authenticity}. Healthcare, journalism, and dating raise further concerns about trust, consent, and anticipated rejection \citep{yoo2025factlabels,toff2025dilemma,barkallah2026dating}.. More detailed disclosures may reduce perceived stigma and increase comfort, but do not by themselves determine whether people are willing to disclose \citep{he2026barriers}. The same disclosure can thus support accountability in one setting while enabling surveillance, misinterpretation, or compelled self-exposure in another. Rather than pointing to a universal label, this evidence frames disclosure as a family of situated governance problems involving stakeholders with diverging needs and unequal power.

These gaps and tensions motivated our FAccT CRAFT workshop, \emph{Who Bears the Cost of Honesty? Co-creating the Future of AI Disclosure}. In this workshop, participants first used scenario-based power mapping to identify stakeholders, relationships, and asymmetries, then created near-future design fiction artifacts to explore more equitable disclosure arrangements and unresolved harms. The organizer-authored scenarios spanned education, work, creative labor, healthcare, politics/journalism, and interpersonal contexts. The archived participant-generated corpus represents four of those domains. The method draws on participatory design's shared ``third spaces'' \citep{muller2003participatory}, design fiction's capacity to make sociotechnical arrangements contestable \citep{blythe2014designfiction}, and workshop synthesis as a way to map an emerging research space \citep{tankelevitch2025tools}. We treat the resulting artifacts not as representative attitudes or evidence of population-level effects, but as situated traces of how participants reasoned about who could demand disclosure, who might benefit, who might bear its costs, and what interventions and residual harms they could imagine.

We make three contributions: (1) a documented workshop method connecting scenario-based power mapping, design fiction, and reflection on residual harms; (2) an artifact-grounded account of how four groups represented disclosure power, possible interventions, and unresolved risks; and (3) the \emph{cost of honesty} as a distributional lens and the \emph{Cost-of-Honesty Stack} as a diagnostic framework, accompanied by provisional design prompts and research directions.

\section{Workshop Design}
The workshop was designed as a field-building and method-generating intervention. Its three goals were to (1) identify stakeholders, needs, harms, and power asymmetries in situated disclosure scenarios; (2) move from critique to the design of alternative disclosure arrangements; and (3) connect researchers and practitioners working on AI disclosure. The in-person program comprised a short introduction, an invited keynote, a 45-minute power-mapping activity, a 45-minute design-fiction studio, and a group share-out. Prof. Aengus Bridgman's keynote brought current AI-disclosure research into conversation with his own experiences and familiar examples, grounding the workshop's inquiry in the practical complexities of how disclosure is encountered in everyday life.

\subsection{Scenario set}
Each group selected a scenario based on its members' interests and used the same scenario across both activities. The six scenario options were intentionally designed as elicitation materials to make power asymmetries and stakeholder tensions concrete. The scenarios and their research motivations were:

\begin{enumerate}[leftmargin=1.6em,itemsep=2pt,parsep=0pt,topsep=3pt]
\item \scenariolabel{scenarioeducation}{Education:} \emph{The AI-Revised Essay}. Language assistance can support multilingual writing, while detector bias and student disclosure requirements make authorship and evaluation contested \citep{ito2023rewriting,liang2023detectors,ye2026norms}. The core tension is assessing authorship without stigmatizing multilingual students.
\item \scenariolabel{scenarioworkplace}{Workplace:} \emph{The AI-Assisted Review}. Evidence of competence penalties and disclosure stigma motivated a scenario that extended these concerns to workplace privacy and accommodations \citep{reif2025social,he2026barriers}. The core tension is seeking transparency without exposing disability-related needs.
\item \scenariolabel{scenariocreative}{Creative work:} \emph{The Watermarked Illustrator}. Contribution type and level shape credit, value, and authenticity judgments in human--AI creative work \citep{he2025credit,schecter2025role,hwang2025authenticity}. The core tension is requiring transparency without erasing human labor and credit.
\item \scenariolabel{scenariohealthcare}{Healthcare:} \emph{The AI Clinical Note-Taker}. Prior work raises questions about purpose, limitations, privacy, fairness, and trust in patient-facing AI information \citep{yoo2025factlabels}. These questions motivated a consent-sensitive scenario. The core tension is improving documentation without compromising patient consent and privacy.
\item \scenariolabel{scenariopolitics}{Politics/journalism:} \emph{The AI-Assisted Election Story}. AI labels can lower audience trust in journalism, motivating scrutiny of newsroom disclosure and accountability \citep{toff2025dilemma}. The core tension is ensuring accountability without turning disclosure into worker surveillance.
\item \scenariolabel{scenariointerpersonal}{Interpersonal relationships:} \emph{The AI-Written Opening Message}. AI-assisted dating self-presentation raises anticipated judgment, authenticity, and informed-consent concerns \citep{barkallah2026dating}. The core tension is supporting authenticity and consent without inviting shame or rejection.

\end{enumerate}

Appendix~\ref{app:workshop-materials} reproduces the full scenario-card text.

\subsection{Keynote talk from Prof. Aengus Bridgman}

Aengus Bridgman's keynote argued that AI disclosure is not neutral. Current evidence shows that simply stating ``this was made with AI'' often reduces trust, even when the content is identical. He emphasized that wording, context, and process transparency matter greatly, and that the most effective practice is to specify what the AI did, who is accountable, and how risks were mitigated, rather than to apply a blunt ``AI-generated'' label. Bridgman also observed that today's disclosure burden is inverted. Honest, less powerful actors such as students and journalists bear the stigma and compliance costs, while major platforms and opaque AI-driven content networks disclose the least. He therefore called for disclosure regimes that focus on powerful actors, emphasize methods over labels, and reorient incentives so that honesty is rewarded rather than punished.

\subsection{Activity 1: Mapping disclosure power}
Groups first identified actors with direct and indirect stakes in their scenario. They categorized actors as primary, secondary, and tertiary according to how directly each actor experienced the effects of AI use and disclosure, then drew and annotated relationships among them. The activity was designed to move discussion beyond a dyad between an AI user and a disclosure recipient. It prompted groups to include institutions, evaluators, platforms, vendors, regulators, and affected communities, and to ask who could impose a requirement, interpret a disclosure, benefit from it, or bear its consequences. The outcome of Activity~1 was a completed power map for each represented scenario.

\subsection{Activity 2: Designing disclosure futures}
Groups then selected one of eight artifact types to co-design in response to their scenario: an interface screen, policy, consent or refusal process, contract or agreement, public label or notice, audit report, community norm or ritual, or training and guidance material. Each artifact-type was introduced through an informational card and served as a starting point for ideation. Groups then designed, sketched, or prototyped an intervention from a near future (circa 2029) intended to establish more equitable and context-aware disclosure norms. They also completed a context card identifying who would benefit, which ``costs of honesty'' the intervention might mitigate, which costs might remain or worsen, and what new harms could emerge. The design and completed context card constituted the outcome of Activity~2. Appendix~\ref{app:workshop-materials} provides the complete Activity~2 instructions, artifact-type cards, and context-card prompts.

\section{Data Corpus and Analysis}

\subsection{Participants, grouping, and facilitation}
Based on a registration form, the workshop included 22 participants, of whom 13 identified as academic researchers or faculty and 9 as undergraduate or graduate students. The data corpus represents outputs from four groups---education, workplace, politics/journalism, and interpersonal relationships---with roughly 3--6 participants each. Creative work and healthcare were not represented among the archived outputs and are therefore excluded from the outcome analysis. Organizers circulated among groups, supported the activities, and convened a final share-out.

\subsection{Collected participant-generated corpus}
The current corpus contains four completed power maps, three completed context cards, and one dating-app interface prototype comprising three screens and a recipient-facing label. The education, interpersonal, and workplace groups completed their context cards. No verbatim transcript or systematic facilitator-note set was collected, so the synthesis below is deliberately grounded in what participants drew, wrote, and designed rather than in recollected discussion.

\subsection{Artifact-centered thematic synthesis}
We conducted an informal artifact-centered thematic synthesis, drawing on thematic analysis to develop and refine patterns across qualitative materials \citep{braun2006using}.

We first transcribed legible handwriting. For each power map, we recorded stakeholder positions, relationships, focus of decision-making power, disclosure obligations, and annotations about needs or risks. For each design-fiction output and context card, we recorded the proposed intervention, intended beneficiaries, information revealed or withheld, user agency, mitigated costs, residual harms, and absent remedies. We then compared patterns within and across domains, attending both to recurrence and to domain-specific differences. The analysis asks four questions: who can demand disclosure, who must account for AI use, who interprets and acts on the disclosure, and what protections, refusals, or remedies are available.

Because the scenarios and artifact-type cards shaped what participants considered, we retain them as elicitation metadata when interpreting each output. We do not, however, code their wording as participant evidence. This makes prompt effects visible while keeping empirical claims grounded in what participants drew, wrote, and designed.

Two organizers independently reviewed a subset of artifact transcriptions and thematic memos. They reconciled ambiguous handwriting by returning to the photographed outputs and comparing interpretations. The review informed refinements to theme names, boundaries, and the supporting evidence attached to each theme.

\section{Workshop Outcomes}
Table~\ref{tab:domainoutputs} summarizes the collected outputs before we draw out patterns represented in more than one artifact. Missing outputs indicate gaps in the collected data corpus, not evidence that no discussion or design work occurred.

\begin{table*}[t]
\footnotesize
\caption{Collected participant-generated outputs by represented domain. Creative work and healthcare were offered as scenarios but not taken up by the participants.}
\label{tab:domainoutputs}
\renewcommand{\arraystretch}{1.12}
\begin{tabular}{p{0.11\linewidth}p{0.20\linewidth}p{0.20\linewidth}p{0.40\linewidth}}
\toprule
\textbf{Domain} & \textbf{Activity 1 output} & \textbf{Activity 2 output} & \textbf{Participant-generated observations visible in the corpus} \\
\midrule
Education & Completed map with a non-native English-speaking student and instructor at the center; an AI tool, English-speaking peers, and an employer or graduate school in the secondary ring; and alumni in the tertiary ring. & Completed context card. & The card names students and instructors as beneficiaries and lists the mitigated costs as stigma and blame around ``using AI,'' ``cheating,'' and ``shortcuts.'' The map depicts peers, the institution, employers, graduate schools, and alumni as stakeholders beyond the student--instructor dyad. \\
Workplace & Completed map with a worker at the center (alongside people with disabilities and people in similar jobs elsewhere); AI-tool providers, customers/clients, an employer or manager, coworkers, and a workplace equality representative in the secondary ring; and society/regulators in the tertiary ring. & Completed context card (provisionally attributed). & Edge labels name the relationships that carry disclosure---``hire or fire / performance review'' (employer to worker), ``responsibility re: work and product quality'' (worker to client), ``inform and use / service needs'' (AI providers to worker), and an equality representative who ``moderates'' the worker--employer tie---with ``social pressure'' spreading through a wider community ``via [a] domino effect.'' The card frames the mitigated cost as the ``privacy concern re: publicizing disability'' and logs unresolved tensions: whether a ``rule-based'' scheme could distinguish ``equalizing'' (accessibility) uses of AI from others, whether it should apply ``only for recognized disabilities,'' and ``how to prevent exploitation.'' \\
Politics/journalism & Completed map connecting a journalist and audience (primary) to an editor, owners/investors, and vendors (secondary), with authority flowing from owners and investors through the editor to the journalist. & No archived Activity~2 output. & The map locates the journalist and audience as directly affected while depicting editorial, ownership, and vendor relationships around them. \\
Interpersonal relationships & Completed map with app users at the center and content moderators just outside them, connecting designers, developers, a product owner, marketing, a CEO, terms-and-conditions/legal, AI platforms, family and friends, and legislators. & Dating-app interface prototype (three screens) and completed context card. & The prototype pairs an AI-use disclosure toggle and an optional ``explain your AI use'' note with a recipient-facing badge that expands into a structured provenance label (``Primarily AI, New content, Human-initiated, Reviewed, ChatGPT v1.0''). The card names both initiator and recipient as beneficiaries, reframes ``disclosure'' as the less shaming ``AI-assisted,'' and notes that stigma or unfair evaluation can persist when a recipient remains judgmental; a risk annotation on the map assigns revenue and compliance risk to the CEO and product owner and reputational risk to family and friends. \\
Creative work & No completed output. & No completed output. & Excluded from artifact-derived outcome claims. \\
Healthcare & No completed output. & No completed output. & Excluded from artifact-derived outcome claims. \\
\bottomrule
\end{tabular}
\end{table*}

\subsection{Distributed impact across multiple actors}
All four power maps placed the AI user---a student, worker, journalist, or dating-app user---among the most directly affected actors. They also depicted instructors, managers, editors, platforms, vendors, owners, clients, and regulators around the disclosure relationship. In the workplace map, edges between the worker and employer are annotated ``performance review'' and ``hire or fire.'' The journalism map connects owners and investors to the editor and journalist, while the education map includes peers, an institution, employers, graduate schools, and alumni. Within these maps, disclosure is represented as a chain involving the person who discloses AI use, the actors who define, interpret, or act on that disclosure, and those affected by its downstream consequences, for example, when interpretations or actions shape later evaluations, opportunities, or institutional decisions.

\subsection{Assistive use as evaluative evidence}
The education and interpersonal context cards connect disclosure to stigma, blame, ``cheating,'' ``shortcuts,'' shame, and unfair evaluation. The workplace card identifies a ``privacy concern re: publicizing disability'' and leaves unresolved whether a rule could distinguish ``equalizing'' uses of AI from other uses without applying ``only for recognized disabilities.'' These artifacts surface a concern that disclosing language, accessibility, drafting, or confidence support could invite judgments about competence, effort, authenticity, or character. This is a risk represented in the artifacts and supported by related evidence on social evaluation and disclosure stigma, not an effect measured in the workshop \citep{reif2025social,he2026barriers}.

\subsection{Contextual disclosure retained interpretive and governance risks}
The interpersonal prototype combined a disclosure toggle with an optional free-text note (``explain your AI use'') and a recipient-facing badge that expanded into a structured label---``Primarily AI, New content, Human-initiated, Reviewed, ChatGPT v1.0''--- modelled after the attribution framework proposed by \citet{he2025credit}. It therefore supplied information about whether AI was involved, the role assigned to it, and human review. The accompanying card proposed the less shaming phrase ``AI-assisted'' and suggested that an explanation could reduce stigma. Although these are design intentions rather than evaluated effects, they illustrate a contextual alternative to a binary label that aligns with evidence on disclosure granularity and reduced stigma \citep{he2026barriers}.

The maps also included actors positioned to seek or interpret information: instructors, employers and clients, editors and audiences, and dating-app recipients. The workplace map placed an equality representative in a role that ``moderates'' the worker--employer relationship, while the interpersonal prototype supplied explanation and recipient-facing provenance. At the same time, the interpersonal card notes that judgment and unfair evaluation could remain, and the workplace card asks ``how to prevent exploitation.'' Explicit appeal, anti-retaliation, deletion, or redress mechanisms are largely absent from the corpus. The artifacts therefore direct attention to governance questions alongside interface choices, including who may collect, retain, interpret, contest, or act on a disclosure.

\section{The Cost-of-Honesty Stack}
To organize these observations, we propose the \emph{Cost-of-Honesty Stack}: a diagnostic framework for locating where AI disclosure is demanded, interpreted, recorded, and enforced within a sociotechnical system. Figure~\ref{fig:stack} presents its five layers and the questions asked at each.

\begin{figure}[t]
\centering
\footnotesize
\fbox{\begin{minipage}{\linewidth}
\centering
\textbf{Governance/public layer}\\
Regulators, publics, professional norms\\[0.2em]
{\itshape Which harms is disclosure meant to prevent? Which groups were included in rule-making? Does policy create legal certainty without social penalties? How are disparate impacts measured?}\\[0.35em]
\rule{\linewidth}{0.4pt}\\[0.35em]
\textbf{Infrastructure layer}\\
AI vendors, provenance systems, watermarking, logs\\[0.2em]
{\itshape What data is captured? Is the system robust? Who can access logs? Which false positives and negatives are expected? How is provenance kept from becoming surveillance?}\\[0.35em]
\rule{\linewidth}{0.4pt}\\[0.35em]
\textbf{Institution/platform layer}\\
Universities, employers, clinics, newsrooms, platforms\\[0.2em]
{\itshape Who wrote the policy? Is it domain-specific? Can people refuse or appeal? Does the institution disclose its own AI use in evaluation and enforcement?}\\[0.35em]
\rule{\linewidth}{0.4pt}\\[0.35em]
\textbf{Evaluation layer}\\
Instructors, managers, clients, editors, audiences, recipients\\[0.2em]
{\itshape What does the evaluator need to know? What assumptions attach to AI use? Can assistance be distinguished from substitution? Are evaluators trained to avoid competence penalties and AI shaming?}\\[0.35em]
\rule{\linewidth}{0.4pt}\\[0.35em]
\textbf{User labor layer}\\
Students, workers, creators, patients, journalists, app users\\[0.2em]
{\itshape What was AI used for? Was the use assistive, generative, accessibility-related, or deceptive? Does disclosure expose disability, language background, workload, or identity?}
\end{minipage}}
\caption{The Cost-of-Honesty Stack. Each layer names the actors through which disclosure passes and the diagnostic questions asked at that layer.}
\Description{A five-layer stack from user labor through evaluation, institution and platform, infrastructure, and governance layers, each annotated with its key diagnostic questions.}
\label{fig:stack}
\end{figure}

At the \emph{user labor layer}, disclosure attaches to the person using AI---the student, worker, creator, journalist, patient, or app user. The questions at this layer ask what AI was used for, whether the use was assistive, generative, accessibility-related, procedural, or deceptive, and whether disclosure exposes sensitive information about disability, language background, workload, or identity.

At the \emph{evaluation layer}, disclosure may be interpreted by someone with the capacity to judge the user or artifact, including instructors, managers, clients, editors, audiences, patients, and interpersonal recipients. Here the diagnostic asks what the evaluator needs to know, what assumptions they attach to AI use, whether they can distinguish assistance from substitution, and whether they are trained to avoid biased competence penalties or AI shaming.

At the \emph{institution/platform layer}, disclosure may become policy. Universities, employers, platforms, clinics, and newsrooms can decide whether disclosure is required, optional, encouraged, prohibited, or ignored. the Stack asks who wrote the policy, whether it is domain-specific, whether it allows refusal, whether it provides mechanisms for appeals and exceptions, and whether the institution discloses its own AI use in evaluation and enforcement.

At the \emph{infrastructure layer}, disclosure may be made machine-readable, logged, detected, watermarked, or audited by AI vendors, provenance standards, detector systems, and content moderation infrastructure. At this layer, the diagnostic asks what data is captured, whether the system is robust, who can access logs, what false positives and false negatives are expected, and how to prevent technical provenance from becoming a surveillance mechanism.

At the \emph{governance/public layer}, disclosure may become a matter of law, public trust, and professional legitimacy. The questions here concern which harms disclosure is meant to prevent, which groups were included in rule-making, whether policy creates legal certainty without imposing social penalties, and how disparate impacts are measured after implementation.

\subsection{Tracing two participant outputs through the Stack}
To show the Stack in use, we read two archived outputs through its layers. These traces are author-generated analytic demonstrations. Where an artifact does not explicitly represent a layer, we use the Stack to pose a question rather than attribute an interpretation to participants.

\emph{Workplace: when accommodation becomes evaluation.} At the \textbf{user-labor layer}, the context card names a ``privacy concern re: publicizing disability.'' At the \textbf{evaluation layer}, the map's ``performance review'' and ``hire or fire'' edges position a manager to interpret the worker's disclosure. At the \textbf{institution/platform layer}, the card asks whether a rule would apply ``only for recognized disabilities,'' raising questions about policy scope and exclusion. The map names AI-tool providers but does not specify what data they retain; at the \textbf{infrastructure layer}, the Stack therefore prompts questions about whether use is recorded and who can access any record. Similarly, the map places society and regulators in its tertiary ring without specifying their actions; at the \textbf{governance/public layer}, the Stack asks who could address the card's concern about exploitation. This reading focuses attention on how evaluation and policy could shape the risk identified in the card.

\emph{Interpersonal: contextual provenance and its limit.} At the \textbf{user-labor layer}, the card reframes ``disclosure'' as the less shaming ``AI-assisted.'' At the \textbf{evaluation layer}, the recipient sees a badge that expands into a structured label (``Primarily AI, New content, Human-initiated, Reviewed''), while the card notes that a ``judgemental'' recipient could still penalize the sender. At the \textbf{institution/platform layer}, the interface determines what the toggle and note reveal, and the map assigns revenue and compliance risk to platform actors. The label names ``ChatGPT v1.0'' but does not specify a provenance system; at the \textbf{infrastructure layer}, the Stack asks whether the label is self-reported or backed by a record. The map includes legislators and content moderators without assigning them a particular intervention; at the \textbf{governance/public layer}, the Stack asks who sets and enforces disclosure rules. The prototype encodes a proposed intervention at the user and evaluation layers while its context card retains the risk of adverse interpretation.

\section{Applying the Stack: Provisional Design Suggestions}
Table~\ref{tab:equitableprinciples} translates recurring questions from the Stack into provisional design suggestions. Their orientation is consistent with data justice's attention to how people are made visible and treated \citep{taylor2017data}, design justice's focus on power and the people most affected by design decisions \citep{costanzachock2020design}, and critiques showing that transparency or technical interventions alone do not ensure accountability or fairness in sociotechnical systems \citep{ananny2018seeing,selbst2019fairness}. Given the scale of our workshop, these suggestions should be interpreted as starting points for adaptation and evaluation rather than validated standards.

\begin{table*}[t]
\footnotesize
\caption{Provisional design suggestions and illustrative domain applications. The examples are not evaluated policy recommendations.}
\label{tab:equitableprinciples}
\renewcommand{\arraystretch}{1.12}
\begin{tabular}{p{0.16\linewidth}p{0.27\linewidth}p{0.48\linewidth}}
\toprule
\textbf{Design suggestion} & \textbf{Question to ask} & \textbf{Illustrative domain applications} \\
\midrule
\textbf{Purpose-limited} & What decision or relationship makes disclosure necessary? & \textbf{Education:} a policy might request only information relevant to the learning objective. \textbf{Journalism:} a notice might focus on AI involvement that materially affects provenance, factual content, or editorial responsibility. \\
\textbf{Least-revealing} & What is the minimum information that serves that purpose? & \textbf{Workplace:} a worker might name a category of assistance without providing full logs or accommodation details. \textbf{Interpersonal:} a label might communicate AI's role without exposing private prompts or motives. \\
\textbf{Differentiate assistance and substitution} & Which forms of AI use are relevant to the judgment being made? & \textbf{Education:} a policy might distinguish language or accessibility support from answer generation. \textbf{Journalism:} a notice might distinguish transcription and copyediting from drafting or synthetic media. Granularity can affect perceived stigma \citep{he2026barriers}. \\
\textbf{Protect refusal and contextualization} & Can a person safely refuse, contest, or explain a disclosure? & \textbf{Workplace:} a process might support contextualization when disclosure could expose disability or precarious status. \textbf{Interpersonal:} an interface might let the sender explain AI's role while supplying consent-relevant information to the recipient. \\
\textbf{Make disclosure reciprocal} & What must institutions, platforms, or evaluators disclose in return? & \textbf{Education:} an institution might state its AI-based grading, detection, and appeal rules. \textbf{Journalism:} a newsroom or platform might report AI use in editing, moderation, or recommendation. \\
\textbf{Support explanation, not confession} & Does the interface enable interpretation without presuming wrongdoing? & \textbf{Interpersonal:} an interface might offer a neutral contextual note and a human-review signal. \textbf{Education:} a course might use a learning-use statement rather than a warning framed around cheating. \\
\textbf{Audit downstream harms} & How will unequal penalties, privacy loss, or misinterpretation be detected and remedied? & \textbf{Education/workplace:} an evaluation could examine uneven enforcement and competence penalties. \textbf{Journalism/interpersonal:} a study could test trust and consent outcomes and examine correction or appeal pathways \citep{reif2025social,toff2025dilemma}. \\
\bottomrule
\end{tabular}
\end{table*}

\section{Research Directions}
The workshop's broader aim was to catalyze a community around equitable AI disclosure. The synthesis suggests three directions for future work.

\textbf{Measuring disclosure harms and benefits.}
Future studies could examine how disclosure granularity, timing, audience, and context relate to trust, competence judgments, authenticity judgments, privacy concerns, and willingness to use AI. Sampling and participatory recruitment could prioritize people who may face disproportionate burdens, such as disabled users, non-native speakers, women, older or junior workers, minoritized creators, students, gig workers, and people in precarious employment.

\textbf{Designing domain-specific disclosure patterns.}
Future work could develop and evaluate domain-specific disclosure patterns rather than assume one universal label. Examples include assistive-use, editorial-provenance, patient-information, student-learning, and creative-attribution patterns, as well as refusal and reciprocal-disclosure processes. Candidate patterns could specify purpose, audience, granularity, retention, risks, and remedy.

\textbf{Building participatory disclosure governance.}
Future work could also examine participatory policy development with affected stakeholders. Power mapping, design fiction, and context cards offer one possible way to surface who benefits, who remains at risk, and what harms a proposed disclosure artifact might create. Whether this method improves policy design requires evaluation beyond the present workshop.

\section{Discussion}
The cost-of-honesty lens reframes AI disclosure from a question of whether information is present to a question of how burdens and protections are allocated. This reframing matters because disclosure is often proposed at moments of institutional anxiety. Universities want to preserve academic integrity, employers want to evaluate productivity, clients want assurance, platforms want trust, newsrooms want credibility, and regulators want public accountability. These are legitimate aims. But if disclosure is implemented as a one-way demand on individuals, it can make vulnerable users carry the moral and evidentiary burden of a broader sociotechnical transition.

The participant-generated outputs illustrate why this matters across the represented domains. The education context card connected disclosure with ``cheating,'' ``shortcuts,'' stigma, and blame. The provisionally attributed workplace card left unresolved how a rule might distinguish ``equalizing'' accessibility uses of AI from others, while the map placed the worker among employers or managers, people with disabilities, an equality representative, clients, coworkers, AI-tool providers, and regulators. The journalism map distributed accountability among the journalist, editor, audience, owners or investors, and vendors. The dating-app prototype paired a structured, expandable label with a way for the sender to contextualize AI assistance and proposed the less shaming phrase ``AI-assisted.'' These examples come from participant-created writing, maps, and design output, rather than from the organizer-authored cards used to prompt the activities.

This synthesis directs attention to governance alongside interface design. Badges, notices, toggles, explanations, and labels can shape interpretation, while policy scope, evaluator practice, data retention, appeal mechanisms, and reciprocal transparency determine how disclosures may be used or contested. A system that records disclosure without a correction or appeal process risks producing evidence that a lower-power user cannot challenge. the Stack therefore includes purpose, proportionality, refusal, contextualization, and remedy as questions for future design and evaluation. These are propositions generated by the synthesis, not effects established by the workshop.

\section{Limitations and Ethics}
This paper synthesizes a small, self-selected workshop corpus. Participants were likely already interested in AI ethics, HCI, FAccT, or disclosure, and their artifacts should not be interpreted as representative of all stakeholders. The artifacts are also speculative and produced under time constraints. Their value lies in surfacing tensions and possible futures, not in proving prevalence.

Handwritten artifacts require careful handling. Raw photos may contain identifiable handwriting, incidental marks, or contextual clues. For future iterations of this two-stage workshop method, we recommend publishing redrawn composite figures and paraphrased examples, unless participants explicitly consent to publication of raw photos. When direct quotations from context cards are used, they should be short, anonymized, and checked against the possibility of participant identification. Any participant list should remain anonymized unless explicit consent was obtained for attribution.

There are also positionality limits. The organizing team includes researchers and practitioners already invested in responsible AI, HCI, design, and AI ethics. That expertise supports synthesis, but it may also shape which harms are most visible. Future work should include more direct participation from affected users, including students, disabled workers, patients, freelancers, non-native speakers, journalists, educators, and people subject to disclosure policies outside research communities.

\section{Conclusion}
AI disclosure is becoming an important mechanism for governing generative AI. It can support accountability, trust, provenance, and consent while also creating risks of suspicion, stigma, and surveillance. Through a synthesis of our FAccT workshop, we use the cost of honesty as a lens on how participants represented these benefits and burdens and develop the Cost-of-Honesty Stack as a diagnostic framework. The documented method, artifact-grounded themes, Stack, design suggestions, and research directions offer starting points for future empirical work, participatory policy development, and domain-specific design without presuming a single universal disclosure standard.

\section*{Acknowledgements}
We thank all workshop participants for the time, ideas, and care they contributed to creating and discussing the artifacts synthesized in this paper. We also thank Aengus Bridgman for an engaging keynote that connected current research on AI disclosure with everyday examples and his own experiences.

\textbf{Generative-AI use disclosure.}
The authors established the paper's goals, methodological framing, evidentiary basis, intended outcomes, and synthesis structure through internal discussion, and drafted the first version. Generative-AI tools were subsequently used interactively to edit and refine later drafts by improving clarity, organization, flow, and concision, identifying passages where claims appeared stronger than the available workshop evidence, suggesting alternative formulations, and supporting searches for relevant literature and checks of citation and reference consistency. The tools were not treated as sources of evidence and did not make final decisions. The authors reviewed and revised all generated text, assessed literature suggestions, checked cited references against source materials, and remain responsible for the paper's interpretations, claims, and final text.

\bibliographystyle{ACM-Reference-Format}
\bibliography{references}

@article{braun2006using,
  author  = {Braun, Virginia and Clarke, Victoria},
  title   = {Using Thematic Analysis in Psychology},
  journal = {Qualitative Research in Psychology},
  year    = {2006},
  volume  = {3},
  number  = {2},
  pages   = {77--101},
  doi     = {10.1191/1478088706qp063oa},
  url     = {https://doi.org/10.1191/1478088706qp063oa}
}

@article{tankelevitch2025tools,
  author        = {Tankelevitch, Lev and Glassman, Elena L. and He, Jessica and Kittur, Aniket and Lee, Mina and Palani, Srishti and Sarkar, Advait and Ramos, Gonzalo and Rogers, Yvonne and Subramonyam, Hari},
  title         = {Understanding, Protecting, and Augmenting Human Cognition with Generative {AI}: A Synthesis of the {CHI} 2025 Tools for Thought Workshop},
  journal       = {arXiv preprint arXiv:2508.21036},
  year          = {2025},
  eprint        = {2508.21036},
  archiveprefix = {arXiv},
  primaryclass  = {cs.HC},
  doi           = {10.48550/arXiv.2508.21036},
  url           = {https://arxiv.org/abs/2508.21036}
}

@misc{acm2025authorship,
  author = {{Association for Computing Machinery}},
  title  = {{ACM} Policy on Authorship},
  year   = {2025},
  note   = {Updated September 16, 2025; accessed July 10, 2026},
  url    = {https://www.acm.org/publications/policies/new-acm-policy-on-authorship}
}

@misc{europeancommission2026transparencycode,
  author = {{European Commission}},
  title  = {Code of Practice on Transparency of {AI}-Generated Content},
  year   = {2026},
  month  = jun,
  note   = {Published June 10, 2026; accessed July 10, 2026},
  url    = {https://digital-strategy.ec.europa.eu/en/policies/code-practice-ai-generated-content}
}

@misc{europeanunion2024aiact,
  author       = {{European Parliament and Council of the European Union}},
  title        = {Regulation ({EU}) 2024/1689 Laying Down Harmonised Rules on Artificial Intelligence ({Artificial Intelligence Act})},
  year         = {2024},
  howpublished = {Official Journal of the European Union, L 2024/1689},
  note         = {Article 50},
  url          = {https://eur-lex.europa.eu/eli/reg/2024/1689/oj}
}

@article{wittenberg2024labeling,
  author  = {Wittenberg, Chloe and Epstein, Ziv and Berinsky, Adam J. and Rand, David G.},
  title   = {Labeling {AI}-Generated Content: Promises, Perils, and Future Directions},
  journal = {An MIT Exploration of Generative AI},
  year    = {2024},
  doi     = {10.21428/e4baedd9.0319e3a6},
  url     = {https://mit-genai.pubpub.org/pub/hu71se89}
}

@inproceedings{elali2024transparent,
  author    = {El Ali, Abdallah and Venkatraj, Karthikeya Puttur and Morosoli, Sophie and Naudts, Laurens and Helberger, Natali and Cesar, Pablo},
  title     = {Transparent {AI} Disclosure Obligations: Who, What, When, Where, Why, How},
  booktitle = {Extended Abstracts of the 2024 CHI Conference on Human Factors in Computing Systems},
  year      = {2024},
  publisher = {Association for Computing Machinery},
  address   = {New York, NY, USA},
  articleno = {342},
  numpages  = {11},
  doi       = {10.1145/3613905.3650750},
  url       = {https://doi.org/10.1145/3613905.3650750}
}

@article{reif2025social,
  author  = {Reif, Jessica A. and Larrick, Richard P. and Soll, Jack B.},
  title   = {Evidence of a Social Evaluation Penalty for Using {AI}},
  journal = {Proceedings of the National Academy of Sciences},
  year    = {2025},
  volume  = {122},
  number  = {19},
  pages   = {e2426766122},
  doi     = {10.1073/pnas.2426766122},
  url     = {https://doi.org/10.1073/pnas.2426766122}
}

@misc{gai2025competence,
  author       = {Gai, Phyliss Jia and Hou, Jiayi and Tu, Yanping},
  title        = {Competence Penalty Is a Barrier to the Adoption of New Technology},
  year         = {2025},
  month        = may,
  howpublished = {SSRN working paper},
  note         = {Version dated May 11, 2025; not peer reviewed},
  doi          = {10.2139/ssrn.5255039},
  url          = {https://ssrn.com/abstract=5255039}
}

@inproceedings{kadoma2025perceptual,
  author    = {Kadoma, Kowe and Metaxa, Dana{\"e} and Naaman, Mor},
  title     = {Generative {AI} and Perceptual Harms: Who's Suspected of Using {LLM}s?},
  booktitle = {Proceedings of the 2025 CHI Conference on Human Factors in Computing Systems},
  year      = {2025},
  publisher = {Association for Computing Machinery},
  address   = {New York, NY, USA},
  pages     = {1--17},
  doi       = {10.1145/3706598.3713897},
  url       = {https://doi.org/10.1145/3706598.3713897}
}

@article{toff2025dilemma,
  author  = {Toff, Benjamin and Simon, Felix M.},
  title   = {{``Or They Could Just Not Use It?''}: The Dilemma of {AI} Disclosure for Audience Trust in News},
  journal = {The International Journal of Press/Politics},
  year    = {2025},
  volume  = {30},
  number  = {4},
  pages   = {881--903},
  doi     = {10.1177/19401612241308697},
  url     = {https://doi.org/10.1177/19401612241308697}
}

@inproceedings{hwang2026freelance,
  author    = {Hwang, Angel Hsing-Chi and Wong, Senya and Chen, Baixiao and He, Jessica and Do, Hyo Jin},
  title     = {{``Better Ask for Forgiveness than Permission''}: Practices and Policies of {AI} Disclosure in Freelance Work},
  booktitle = {Proceedings of the 2026 CHI Conference on Human Factors in Computing Systems},
  year      = {2026},
  publisher = {Association for Computing Machinery},
  address   = {New York, NY, USA},
  pages     = {1--16},
  doi       = {10.1145/3772318.3791920},
  url       = {https://doi.org/10.1145/3772318.3791920}
}

@inproceedings{barkallah2026dating,
  author    = {Barkallah, Meryem and Zytko, Douglas},
  title     = {{``I Wanted Them to Think That I Wrote That''}: {AI}-Generated Self-Presentation on Dating Apps and Implications of Non-Disclosure on Informed Consent},
  booktitle = {Proceedings of the 2026 CHI Conference on Human Factors in Computing Systems},
  year      = {2026},
  publisher = {Association for Computing Machinery},
  address   = {New York, NY, USA},
  articleno = {112},
  numpages  = {18},
  doi       = {10.1145/3772318.3791593},
  url       = {https://doi.org/10.1145/3772318.3791593}
}

@inproceedings{he2026barriers,
  author    = {He, Jessica and Akella, Avinash and Do, Hyo Jin and Weisz, Justin D.},
  title     = {Overcoming Barriers to {AI} Disclosure: The Role of Granularity and Sociocultural Enablers in Reducing Perceived Stigmas},
  booktitle = {Proceedings of the 2026 ACM Conference on Fairness, Accountability, and Transparency},
  year      = {2026},
  publisher = {Association for Computing Machinery},
  address   = {New York, NY, USA},
  pages     = {2117--2143},
  doi       = {10.1145/3805689.3806737},
  url       = {https://doi.org/10.1145/3805689.3806737}
}

@inproceedings{he2025credit,
  author    = {He, Jessica and Houde, Stephanie and Weisz, Justin D.},
  title     = {Which Contributions Deserve Credit? Perceptions of Attribution in Human-{AI} Co-Creation},
  booktitle = {Proceedings of the 2025 CHI Conference on Human Factors in Computing Systems},
  year      = {2025},
  publisher = {Association for Computing Machinery},
  address   = {New York, NY, USA},
  articleno = {540},
  numpages  = {18},
  doi       = {10.1145/3706598.3713522},
  url       = {https://doi.org/10.1145/3706598.3713522}
}

@article{hwang2025authenticity,
  author  = {Hwang, Angel Hsing-Chi and Liao, Q. Vera and Blodgett, Su Lin and Olteanu, Alexandra and Trischler, Adam},
  title   = {{``It Was 80\% Me, 20\% {AI}''}: Seeking Authenticity in Co-Writing with Large Language Models},
  journal = {Proceedings of the ACM on Human-Computer Interaction},
  year    = {2025},
  volume  = {9},
  number  = {2},
  pages   = {1--41},
  doi     = {10.1145/3711020},
  url     = {https://doi.org/10.1145/3711020}
}

@inproceedings{ito2023rewriting,
  author    = {Ito, Takumi and Yamashita, Naomi and Kuribayashi, Tatsuki and Hidaka, Masatoshi and Suzuki, Jun and Gao, Ge and Jamieson, Jack and Inui, Kentaro},
  title     = {Use of an {AI}-Powered Rewriting Support Software in Context with Other Tools: A Study of Non-Native English Speakers},
  booktitle = {Proceedings of the 36th Annual ACM Symposium on User Interface Software and Technology},
  year      = {2023},
  publisher = {Association for Computing Machinery},
  address   = {New York, NY, USA},
  articleno = {45},
  numpages  = {13},
  doi       = {10.1145/3586183.3606810},
  url       = {https://doi.org/10.1145/3586183.3606810}
}

@misc{ye2026norms,
  author        = {Ye, Runlong and Huang, Oliver and He, Jessica and Liut, Michael},
  title         = {Exploring Emerging Norms of {AI} Attribution and Disclosure in Programming Education},
  year          = {2026},
  howpublished  = {arXiv preprint arXiv:2602.04023},
  eprint        = {2602.04023},
  archiveprefix = {arXiv},
  primaryclass  = {cs.HC},
  doi           = {10.48550/arXiv.2602.04023},
  url           = {https://arxiv.org/abs/2602.04023}
}

@inproceedings{schecter2025role,
  author    = {Schecter, Aaron and Richardson, Benjamin},
  title     = {How the Role of Generative {AI} Shapes Perceptions of Value in Human-{AI} Collaborative Work},
  booktitle = {Proceedings of the 2025 CHI Conference on Human Factors in Computing Systems},
  year      = {2025},
  publisher = {Association for Computing Machinery},
  address   = {New York, NY, USA},
  numpages  = {15},
  doi       = {10.1145/3706598.3713946},
  url       = {https://doi.org/10.1145/3706598.3713946}
}

@inproceedings{yoo2025factlabels,
  author    = {Yoo, Dong Whi and Stroud, Austin M. and Zhu, Xuan and Miller, Jennifer E. and Barry, Barbara},
  title     = {Toward Patient-Centered {AI} Fact Labels: Leveraging Extrinsic Trust Cues},
  booktitle = {Proceedings of the 2025 ACM Designing Interactive Systems Conference},
  year      = {2025},
  publisher = {Association for Computing Machinery},
  address   = {New York, NY, USA},
  pages     = {676--690},
  doi       = {10.1145/3715336.3735758},
  url       = {https://doi.org/10.1145/3715336.3735758}
}

@article{liang2023detectors,
  author  = {Liang, Weixin and Yuksekgonul, Mert and Mao, Yining and Wu, Eric and Zou, James},
  title   = {{GPT} Detectors Are Biased Against Non-Native English Writers},
  journal = {Patterns},
  year    = {2023},
  volume  = {4},
  number  = {7},
  pages   = {100779},
  doi     = {10.1016/j.patter.2023.100779},
  url     = {https://doi.org/10.1016/j.patter.2023.100779}
}

@article{taylor2017data,
  author  = {Taylor, Linnet},
  title   = {What Is Data Justice? The Case for Connecting Digital Rights and Freedoms Globally},
  journal = {Big Data \& Society},
  year    = {2017},
  volume  = {4},
  number  = {2},
  pages   = {1--14},
  doi     = {10.1177/2053951717736335},
  url     = {https://doi.org/10.1177/2053951717736335}
}

@book{costanzachock2020design,
  author    = {Costanza-Chock, Sasha},
  title     = {Design Justice: Community-Led Practices to Build the Worlds We Need},
  publisher = {MIT Press},
  address   = {Cambridge, MA},
  year      = {2020},
  doi       = {10.7551/mitpress/12255.001.0001},
  url       = {https://doi.org/10.7551/mitpress/12255.001.0001}
}

@inproceedings{selbst2019fairness,
  author    = {Selbst, Andrew D. and Boyd, Danah and Friedler, Sorelle A. and Venkatasubramanian, Suresh and Vertesi, Janet},
  title     = {Fairness and Abstraction in Sociotechnical Systems},
  booktitle = {Proceedings of the Conference on Fairness, Accountability, and Transparency},
  year      = {2019},
  publisher = {Association for Computing Machinery},
  address   = {New York, NY, USA},
  pages     = {59--68},
  doi       = {10.1145/3287560.3287598},
  url       = {https://doi.org/10.1145/3287560.3287598}
}

@incollection{muller2003participatory,
  author    = {Muller, Michael J.},
  title     = {Participatory Design: The Third Space in {HCI}},
  booktitle = {The Human-Computer Interaction Handbook: Fundamentals, Evolving Technologies, and Emerging Applications},
  editor    = {Jacko, Julie A. and Sears, Andrew},
  publisher = {Lawrence Erlbaum Associates},
  address   = {Mahwah, NJ},
  year      = {2003},
  pages     = {1051--1068}
}

@inproceedings{blythe2014designfiction,
  author    = {Blythe, Mark},
  title     = {Research through Design Fiction: Narrative in Real and Imaginary Abstracts},
  booktitle = {Proceedings of the SIGCHI Conference on Human Factors in Computing Systems},
  year      = {2014},
  publisher = {Association for Computing Machinery},
  address   = {New York, NY, USA},
  pages     = {703--712},
  doi       = {10.1145/2556288.2557098},
  url       = {https://doi.org/10.1145/2556288.2557098}
}

@article{ananny2018seeing,
  title={Seeing without knowing: Limitations of the transparency ideal and its application to algorithmic accountability},
  author={Ananny, Mike and Crawford, Kate},
  journal={new media \& society},
  volume={20},
  number={3},
  pages={973--989},
  year={2018},
  publisher={SAGE Publications Sage UK: London, England}
}

\appendix
\section{Workshop Activity Materials}
\label{app:workshop-materials}

This appendix reproduces the organizer-authored wording used in the workshop so that readers can inspect, adapt, or reuse the elicitation materials. The scenario cards, artifact-type cards, instructions, and blank context-card prompts below are workshop inputs, not participant-generated findings. Their visual formatting has been condensed for print, but the substantive text is reproduced from the activity materials.

\subsection{Activity 2 Instructions}

\paragraph{How to Use the Cards.}
\emph{Disclosure Design Fiction Studio - Scenario + Artifact + Context}

\begin{enumerate}[leftmargin=*,label=\arabic*.]
    \item \textbf{Choose one Scenario Card.} This gives your group a concrete disclosure tension and stakeholders.
    \item \textbf{Choose one Artifact Type Card.} This gives your group a form to design: an interface, policy, ritual, label, report, process, or guide.
    \item \textbf{Complete the sentence.} ``We are designing a 2029 \underline{\hspace{1.2cm}} for the \underline{\hspace{1.2cm}} scenario that helps \underline{\hspace{1.2cm}} without \underline{\hspace{1.2cm}}.''
    \item \textbf{Make the artifact and context card.} Sketch, write, storyboard, or prototype the artifact, along with a completed context card.
\end{enumerate}

\noindent\textbf{Share the artifact and context card.} Show the artifact your group made and use the context card to explain who it helps, which risks remain, and what protections would be needed in practice.

\subsection{Context Card Prompts}

\noindent\emph{Who Benefits, Who Remains at Risk?}

\noindent Complete this after your group sketches the 2029 artifact. Use short phrases and leave space for details you want to share.

\paragraph{Who benefits?}
Who gains trust, clarity, credit, protection, control, time, money, or legitimacy?

\paragraph{Which ``costs of honesty'' are mitigated?}
What stigma, surveillance, blame, privacy loss, unfair evaluation, or lost credit is reduced?

\paragraph{Which ``costs of honesty'' are not mitigated, or get worse?}
Who still carries risk? Does the artifact shift costs onto someone with less power?

\paragraph{What additional harms might emerge, or remain unaddressed?}
What new risks, exclusions, misunderstandings, or enforcement problems could appear?

\paragraph{Context sentence.}
``This artifact helps \underline{\hspace{1.4cm}} by \underline{\hspace{1.4cm}} but still needs \underline{\hspace{1.4cm}}.''

\begin{table*}[t]
\footnotesize
\caption{Complete organizer-authored scenario cards. The first column identifies the domain and card title; the second reproduces the full card text.}
\label{tab:full-scenario-cards}
\setlength{\tabcolsep}{5pt}
\renewcommand{\arraystretch}{1.08}
\begin{tabular}{>{\raggedright\arraybackslash}p{0.18\linewidth}p{0.76\linewidth}}
\toprule
\textbf{Scenario and title} & \textbf{Full card text} \\
\midrule
\scenariolabel{scenarioeducation}{Education}\par\emph{The AI-Revised Essay}
&
\textbf{Situation.} A university student who is not a native English speaker uses AI to review and revise a first writing assignment.\newline
\textbf{Disclosure pressure.} The course policy requires detailed disclosure of how AI was used and why.\newline
\textbf{Core tension.} Instructors want to assess learning and authorship; the student fears their writing will be judged as less competent or less authentic.\newline
\textbf{Watch for.} Language bias; cheating suspicion; reduced trust.\newline
\textbf{Scenario focus.} Consider how language support, authorship expectations, and stigma shape whether disclosure feels safe. \\
\midrule
\scenariolabel{scenarioworkplace}{Workplace}\par\emph{The AI-Assisted Review}
&
\textbf{Situation.} A worker with dyslexia uses AI-powered text-to-speech, speech-to-text, and autocomplete to read and write long documents.\newline
\textbf{Disclosure pressure.} The company performance review asks employees to report how much they use AI day to day.\newline
\textbf{Core tension.} The employer wants transparency; the worker may be forced to reveal disability-related support needs.\newline
\textbf{Watch for.} Ableism; privacy loss; competence penalty.\newline
\textbf{Scenario focus.} Consider when transparency about AI use can force disclosure of disability-related support needs. \\
\midrule
\scenariolabel{scenariocreative}{Creative Work}\par\emph{The Watermarked Illustrator}
&
\textbf{Situation.} A freelance illustrator uses AI during ideation and sketching, then produces the final work by hand.\newline
\textbf{Disclosure pressure.} A major client requires an AI watermark on any work involving AI at any stage.\newline
\textbf{Core tension.} The client wants transparency; the illustrator worries the label will erase their labor and lower the perceived value of the work.\newline
\textbf{Watch for.} Loss of credit; lower pay; flattened authorship.\newline
\textbf{Scenario focus.} Consider how labels can flatten human labor, credit, and value when AI is part of a creative process. \\
\midrule
\scenariolabel{scenariohealthcare}{Healthcare}\par\emph{The AI Clinical Note-Taker}
&
\textbf{Situation.} During an appointment with a transgender patient, a resident physician uses an AI device that records and summarizes the conversation.\newline
\textbf{Disclosure pressure.} The clinic has no clear policy for explaining the AI tool, although patients may be able to opt out.\newline
\textbf{Core tension.} Clinicians want efficient documentation; the patient may fear privacy loss, misgendering, surveillance, or future misuse of sensitive data.\newline
\textbf{Watch for.} Medical mistrust; identity harm; unclear consent.\newline
\textbf{Scenario focus.} Consider how AI use is explained, negotiated, and refused in sensitive care settings. \\
\midrule
\scenariolabel{scenariopolitics}{Politics and Journalism}\par\emph{The AI-Assisted Election Story}
&
\textbf{Situation.} A junior journalist under time pressure uses AI to retrieve and summarize statistics about public perceptions of candidates.\newline
\textbf{Disclosure pressure.} The newsroom requires disclosure of substantive AI use, and editors can request full prompt histories.\newline
\textbf{Core tension.} The newsroom wants audience trust; the journalist may face surveillance, blame, or punishment for routine research assistance.\newline
\textbf{Watch for.} Worker surveillance; public distrust; unfair blame.\newline
\textbf{Scenario focus.} Consider how accountability requirements can become worker surveillance or unfair blame. \\
\midrule
\scenariolabel{scenariointerpersonal}{Interpersonal Relationships}\par\emph{The AI-Written Opening Message}
&
\textbf{Situation.} A dating app user drafts conversation starters with AI to sound more confident and attentive to another person's interests.\newline
\textbf{Disclosure pressure.} There is no formal rule, but people disagree about whether AI-assisted self-presentation should be disclosed.\newline
\textbf{Core tension.} One person may want authenticity and informed consent; the AI user may fear embarrassment, rejection, or judgment.\newline
\textbf{Watch for.} Shame; loss of trust; unclear consent.\newline
\textbf{Scenario focus.} Consider how AI-assisted self-presentation affects authenticity, consent, shame, and trust. \\
\bottomrule
\end{tabular}
\end{table*}

\clearpage

\begin{table*}[t]
\footnotesize
\caption{Complete organizer-authored artifact-type cards. The first column identifies the task or artifact form; the second reproduces the full card text.}
\label{tab:full-artifact-cards}
\setlength{\tabcolsep}{5pt}
\renewcommand{\arraystretch}{1.08}
\begin{tabular}{>{\raggedright\arraybackslash}p{0.18\linewidth}p{0.76\linewidth}}
\toprule
\textbf{Task or artifact type} & \textbf{Full card text} \\
\midrule
\tasklabel{Interface Screen}
&
\textbf{What you make.} A screen, flow, dashboard, or form from a future system.\newline
\textbf{Best for.} Showing what information is collected, who sees it, and what choices users have.\newline
\textbf{Include.} Permissions; visibility; user choices; protections.\newline
\textbf{Starting prompt.} Sketch the first screen someone sees and the decisions they can make. \\
\midrule
\tasklabel{Policy}
&
\textbf{What you make.} A plain-language policy that explains when disclosure is required, optional, protected, or unnecessary.\newline
\textbf{Best for.} Clarifying expectations, responsibilities, exceptions, and appeal rights.\newline
\textbf{Include.} Requirements; exceptions; rights; appeal path.\newline
\textbf{Starting prompt.} Start in broadstrokes, outlining the key policy points a real person would need to understand. \\
\midrule
\tasklabel{Consent or Refusal Process}
&
\textbf{What you make.} A step-by-step process that lets someone consent, refuse, negotiate, or withhold disclosure safely.\newline
\textbf{Best for.} Designing a right to refusal, appeal, contextual disclosure, or protected non-disclosure.\newline
\textbf{Include.} Before; during; after; what refusal triggers.\newline
\textbf{Starting prompt.} Map what happens before, during, and after someone consents or refuses. \\
\midrule
\tasklabel{Contract or Agreement}
&
\textbf{What you make.} A future agreement between people, platforms, clients, institutions, or communities.\newline
\textbf{Best for.} Showing how power, credit, risk, and disclosure obligations are negotiated.\newline
\textbf{Include.} Must disclose; need not disclose; misuse protections.\newline
\textbf{Starting prompt.} Write 3 to 5 clauses that make the power relationship visible. \\
\midrule
\tasklabel{Public Label or Notice}
&
\textbf{What you make.} A label, watermark, audience note, footnote, profile marker, or public explanation.\newline
\textbf{Best for.} Exploring what audiences deserve to know and what details may be harmful to expose.\newline
\textbf{Include.} What is revealed; what is hidden; why.\newline
\textbf{Starting prompt.} Design the label, then explain what it deliberately does not reveal. \\
\midrule
\tasklabel{Audit or Accountability Report}
&
\textbf{What you make.} A one-page report reviewing whether a disclosure system is fair in practice.\newline
\textbf{Best for.} Looking beyond individual disclosure toward institutional responsibility.\newline
\textbf{Include.} Who benefited; who was harmed; what must change.\newline
\textbf{Starting prompt.} Create a report card for the disclosure system itself. \\
\midrule
\tasklabel{Community Norm or Ritual}
&
\textbf{What you make.} A social practice, conversation ritual, group agreement, or shared expectation.\newline
\textbf{Best for.} Designing norms, not just tools, and changing how people talk about AI use.\newline
\textbf{Include.} Who speaks; what is made safe; what is off-limits.\newline
\textbf{Starting prompt.} Storyboard how the conversation happens in real life. \\
\midrule
\tasklabel{Training or Guidance}
&
\textbf{What you make.} A checklist, mini-guide, training handout, or evaluator aid.\newline
\textbf{Best for.} Reducing stigma by changing how instructors, managers, editors, clinicians, or audiences interpret AI use.\newline
\textbf{Include.} Do not assume; ask instead; fair evaluation criteria.\newline
\textbf{Starting prompt.} Create a short guide for the person who might misuse disclosure. \\
\bottomrule
\end{tabular}
\end{table*}

\end{document}